\documentstyle[preprint,aps,epsf]{revtex} 
\begin{document}

\draft

\title{Ferromagnetic correlations in quasi-one-dimensional conducting
channels}

\author{Boris Spivak}

\address{Physics Department, University of Washington, Seattle, WA 98195,
USA}

\author{Fei Zhou}

\address{Physics Department, Princeton University, Princeton, NJ 08544,
USA}

\maketitle

\begin{abstract}
We propose a model which explains the experimental observation of
spontaneous spin polarization of conducting electrons in
quasi-one-dimensional
 AlGaAs/GaAs channels \cite{tom,tom1}. We show that a ferromagnetic
order
is a generic property of a quasi-one- dimensional conducting channel
embedded in a Wigner crystal. 
We also discuss gate voltage, magnetic
field and
temperature
dependences of the channel's conductance.
\end{abstract}

\newpage

Recently both theories and experiments on conductances of quasi-one
dimensional
conducting channels attracted a lot of attention (See for example 
\cite{tren,wees,tom,tom1,binakk}).
Quasi-one-dimensional 
confining potential has been made by applying split gates
 to two-dimensional 
electron gas in AlGaAs/GaAs heterostructures (See the insert in Fig.1). In
this case electron
motion is
quantized in $x$ direction.
If the confining potential changes adiabatically in
 "y" direction,
we can neglect backscatterings of electrons incident to the
 channel from the two-dimensional reservoirs. For
noninteracting electrons at zero temperature and in the absence of an
external magnetic field $H$, the conductance of the system is quantized
\begin{equation}
G=\frac{2e^{2}}{h}k_{1}.
\end{equation} 
Here $k_{1}=0,1,2...$ is an integer associated with number of quantized
 levels in $x$ direction, which are filled with electrons. The prefactor $2$
in
Eq.1
reflects the 
spin degeneracy of electron states at $H=0$. 
In experiments \cite{tren,wees,tom,tom1}, the electron concentration in
two-dimensional
semiconductor structures was
controlled by a gate voltage $V_{g}$.  Eq.1 would manifest itself in the
existence of plateaus in the $V_{g}$-dependence of $G(V_g)$ quantized in
units
$\frac{2e^{2}}{h}$.
When the concentration of electrons in the channel is high,
or  $k_{1}\gg
1$,  Eq.1 is in
very good agreement with experimental data \cite{tren,wees,tom,tom1}.
However, at small concentration of electrons, when $k_{1}\sim 1$, experimental
data  \cite{tom,tom1} deviate from Eq.1 dramatically.
The following anomalies have been observed:

a). At small temperature $T$, in addition to plateaus at
$G=\frac{2e^2}{h},
2\frac{2e^2}{h}$, the dependence $G(V_{g})$ exhibits "smeared
plateaus" at
$G\sim 0.7\frac{2e^2}{h}$ and $G\sim 1.7\frac{2e^2}{h}$. 

Furthermore, in the noninteracting electron picture, an external
magnetic field $H$ parallel to the film lifts the spin degeneracy and
each
plateau splits into two plateaus quantized in units $\frac{e^{2}}{h}$.
Again, this is in agreement with experiments at large $k_{1}$.
However, the plateaus at $G=0.7\frac{2e^2}{h}, \frac{2e^2}{h}
 1.7\frac{2e^2}{h}, 2\frac{2e^2}{h}$ 
 do not split in the external magnetic field \cite{tom,tom1}. It has been
noted \cite{tom,tom1}
that the $H$ dependence of positions of these plateaus can be
explained if one assumes a ferromagnetic electron ground
state in the channel at $H=0$.

b). The first and third plateau ($G\sim 0.7\frac{2e^{2}}{h}, 1,7
 \frac{2e^{2}}{h}$) are more smeared
than
the second
and fourth plateau ($G=\frac{2e^{2}}{h}, 2\frac{2e^{2}}{h}$).

 c) The plateaus at $G=\frac{2e^{2}}{h}, 2\frac{2e^2}{h}$
become weaker as T increases, while
 the plateaus at $G\sim 0.7\frac{2e^{2}}{h}, 1.7\frac{2e^{2}}{h}$
become stronger.

Existing theories of one-dimensional interacting
electron gas predict no spin polarization in the ground state 
and can not explain the measured dependence $G(V_{g}, H,T)$ when the 
concentration of electrons in the channel is small.

In this article we propose a qualitative explanation of experimentally
 observed anomalous features.
Consider the confining potential $V(x,V_{g})=V_{0}(V_{g})(1+
\frac{x^{2}}{A^{2}(V_{g})})$
as shown in  Fig.1. 
If the potential $V(x)$ is shallow (the value of $A$ is big) and
the concentration of electrons in the channel is small, the ground state
of the electron system is a Wigner crystal. In Fig.1 positions of
 localized electrons in the Wigner crystal are indicated by solid dots. We
will assume 
that the Wigner crystal is pinned and the conductance of the system is
determined
by hoppings of point defects in the crystal (interstitial electrons or
vacancies), concentration of which is exponentially small. 
 The activation energy $E^{i}(x)$ of interstitial electrons is nonuniform
and
 has a minimum $E^{i}_{m}(V_{0})=E^{i}(x=0)$ in the middle of the
channel. 

At a critical value $V_{0}=V_{0c}$, melting of Wigner crystal occurs 
near $x=0$ and a quasi-one dimensional conducting
channel is formed. Below we consider two scenarios where 
creation of the channel corresponds to a first
and second
order phase transition. In both scenarios the channel is ferromagnetic.

\section{ Second order Lifshitz phase transition}

Let us assume that at the critical value $V_{0}=V_{c0}$, for the first
time, the
activation energy of the interstitials becomes zero in the middle of
the channel, i.e. $E^{i}_{m}(V_{0}=V_{0c})=0$.
This can be considered as a second order Lifshitz phase
transition. The interstitial electrons are indicated in Fig.1 by
the symbol $"*"$.
This transition is
a precursor
of melting of a crystal into a liquid.
 The existence of zero point motion
defects in
the middle of the channel is a nonuniform
analogy of the model of supersolid proposed by A.F.Andreev and I.M
Lifshitz
 in the context of quantum melting of helium crystals \cite{andlif}. 
Such a state has never been observed in bulk helium crystals.
Latter it was noted \cite{andpa} that conditions for existence of
such a state are 
less restrictive at a crystal-liquid surface, where the amplitude of
quantum fluctuations of
the lattice sites is larger. This leads
to a quantum rough state of the surface, where zero point motion steps
on the surface exist in the ground state, which, in turn, leads to the
existence of surface 
crystallization waves at $T=0$ \cite{andpa}. Crystallization waves at
small $T$, have been observed \cite{paba}, while 
existence of solid-liquid helium quantum rough surfaces in d=3 case
 is still an open question \cite{fish,ior}. In two dimensional case 
the existence of quantum rough surfaces is more plausible than in 
the three dimensional case because the amplitude of quantum fluctuations
of particles at the surface is larger.

To begin, we first consider the case when all spins are polarized in
$\downarrow$ direction, for example,
by an external magnetic field parallel to the film. In this case the
interstitial's energy
spectrum in the quasi-one dimensional channel has a form 
\begin{equation}
\epsilon_{\downarrow}(k_{x},p_{y})=\epsilon_{0}(k_{x}+\frac{1}{2})+
\Delta_{\downarrow}\cos(ap_{y}).
\end{equation}
Here $k_{x}=1,2..$ is an integer corresponding to quantization of the wave
function in $x$ direction, $p_{y}$ is a momentum in $y$ direction, 
$\epsilon_{0}=\frac{\hbar^2}{2m^{*}L^2}$, $L=
(\frac{\hbar A}{2\sqrt{2 eV_{0}m^{*}}})^{\frac{1}{2}}$ is a characteristic
dimension of the wave function in $x$ direction,
$m^{*}\sim   
\frac{\hbar^{2}}{\Delta_{\downarrow}a^{2}}$ is the
effective mass of the interstitial, $\Delta_{\downarrow}$ is the
 band width for interstitials with spin $\downarrow$ direction, and
$a=\frac{1}{\sqrt{n}}$ and $n$ are interelectron distance and electron
concentration in the Wigner crystal respectively.
 Band structures for
interstitials moving in a polarized  Wigner crystal in the case
$\epsilon_{0}>\Delta_{\downarrow}$ are shown in Fig.2.

In the absence of an external magnetic field $H=0$, the interstitial
electrons wave
 functions and spectrum inside the conducting channel depend crucially on 
the magnetic structure of the Wigner crystal. The magnetic structure of 
a uniform Wigner crystal is not known (See a discussion in \cite{chak}).
We
will
assume that the magnetic exchange
energy between electrons $J$ has an antiferromagnetic sign and the ground
state
corresponds to either an antiferromagnetic state or a resonating valence
bonds (RVB) state \cite{rvb}. We will
show that in both cases the conducting
channel should be ferromagnetic. 
It has been noted \cite{her,ferow} that in some range of concentrations the
ferromagnetic state of the Wigner crystal is also possible. In this case, 
our results are
valid at relatively high temperatures $T>J$. We believe that in 
experiments \cite{tom,tom1} this inequality took place because in the
Wigner crystal $J\sim 1mK$ \cite{ferow}.

It has been known for a long time
\cite{nag,thouless,nagae,nagae1,brink,and} that
a point defect
embedded in
a uniform antiferromagnetic quantum crystal is surrounded by
a ferromagnetic region with a radius 
\begin{equation}
R\approx a(\frac{\Delta_{\downarrow}}{J})^{1/4}\gg a.
\end{equation}
We assume that $\Delta_{\downarrow}\gg J$.
The reason for the existence of the magnetic polaron is  that
 $\Delta_{\downarrow}$ is much larger than
the energy band width
of the defect embedded in an antiferromagnetic crystal. A competition
 between
negative delocalization
energy of the point defect inside the ferromagnetic polaron ( of
 order of
$\frac{\hbar^{2}}{2R^{2}m^{*}}$) and the positive spin polarization energy
of
the magnetic polaron,
(of order of $J(\frac{R}{a})^{2}$) leads to Eq.3.

If $T> J$ one should substitute $J$ with $T$ in Eq.3.

Thus there are
two characteristic lengths in the problem, $L$ and $R$.
Let us consider the case $L<R$ when the magnetic polarons
have sizes $R$ and $L$ in $y$ and $x$ direction respectively.
 If $N$, the
one-dimensional concentration of the interstitial electrons in
the channel, is small ($N\ll R^{-1}$), the magnetic polarons do not
overlap;
the
magnetization $\bbox{M}(y)$ is nonuniform; the channel
conducts poorly and the conductance of the
channel increases gradually with $N$ (or with $V_{g}$).  At $N\gg R^{-1}$,
the magnetic polarons overlap and form a conducting magnetically polarized
channel.
The width of the magnetically polarized region in this case is of order
of $L$. At this point the concentration of interstitials in the channel is
 small and, due to
exchange spin splitting, they have spins $\downarrow$.
The conductance of the channel exhibits the first plateau at
$G=\frac{e^2}{h}$.

The situation arising at a high concentration of interstitials $N$ depends
on a
ratio between $\epsilon_{0}$ and $\Delta_{\downarrow}$. If
 $\epsilon_{0}\gg \Delta_{\downarrow}$, further
increasing of $N$
will lead to filling of the band $k=1,\uparrow$. Since $\Delta_{\uparrow}
< \Delta_{\downarrow}$, the kinetic
energy of
 interstitials with
spins $\uparrow$ is small (See Fig.2).  Here
$\Delta_{\uparrow}$
 is the width
of the band with $k=1,\uparrow$. The width of the
magnetic stripe does not
increase as the band $k=1,\uparrow$ is being filled and
interstitials in the band $k=1,\uparrow$ move in the uniform exchange
potential associated  with the uniform magnetization $\bbox{M}$ inside the
channel. Therefore the conductance exhibits the second plateau at
$G=\frac{2e^{2}}{h}$, which is not smeared.

 At last, let us discuss the temperature dependence of the first
plateau, which corresponds to the band
$k=1,\downarrow$. Suppose  that $(V_{0}-V_{0c})<0$ and $|V_{0}-V_{0c}|$ is
 small and, consequently, the
interstitial activation
energy $E^{i}_{m} >0$ is also small. In this situation 
interstitials thermally activated in the channel
increase the magnetic moment ${\bf M}$ and make the
distribution of $\bbox{M}(z)$ more uniform. As a result, the first plateau
at $G\sim \frac{e^{2}}{h}$ becomes stronger as temperature
increases. Since $E_{m}^{i}$ is small,
 one can neglect thermal fluctuations of the Wigner crystal lattice.
 This is
different from solid $^{3}He$ case where the activation energy of
point defects is large. Therefore the crystal melts before
point defects contribute significantly to magnetic properties of the
sample.

The presented above model explains the following experimental facts.

a).It predicts the ferromagnetic ground state of the conducting 
channel and the existence of an additional step in the $V_g$-dependence
$G(V_{g}$, H=0) at $G=\frac{e^{2}}{h}$.
 
b).It predicts that the first plateau in
the $V_g$-dependence of $G(V_{g})$ is more smeared than the
second one.

 c).It explains why the
first plateau at $G\sim \frac{e^{2}}{h}$ becomes stronger as $T$
increases.

d).It also follows from the presented above picture that at small
 concentration of
electrons in the channel when $G\ll \frac{e^{2}}{h}$, and at small
$T$, the system should exhibit giant negative magnetoresistance
in a parallel magnetic field. This is because the radius of
the
magnetic polaron $R(H)$ increases with $H$. Obviously, this effect takes place
also in the variable range hopping regime \cite{spiv}.

At last, let us mention difficulties of the presented above model.

 1). Experimental data \cite{tom,tom1} exhibit the plateau at
$0.7\frac{2e^{2}}{h}$.
It is not clear to us whether it is possible to attribute the difference 
between $0.7\frac{2e^{2}}{h}$ and $\frac{e^{2}}{h}$ to
smearing of the first plateau.

 2). There are experimental indications
that the first and the third plateau disappear at $T=0$
\cite{kris,kris1}. \footnote{We are grateful
to
D.Khmelnitskii for turning our attention to this difficulty.}
 To this
end, we would
like to mention that the
filling of the first band, as $V_{g}$ changes, is
a complicated process which involves overlapping of the magnetic
polarons. Therefore, even in the
approximation when there is no direct interaction between the
interstitials, there is an indirect
long range many-body effective interaction between the
interstitials, which is mediated by spin degrees of freedom of the Wigner
crystal.
 As a result, if $N$ is fixed, the spatial phase
separation
 of the interstitials takes place at
relatively small $N$ \cite{nagae1}. 
Recently this phenomenon has
been
reconsidered in the context of HTC \cite{kiv}. The apparent
differences
between \cite{kiv} and the situation discussed here are: a)the
interstitials are neutral objects
because they are
screened by
Wigner crystal lattice and b) in experiments it is the chemical potential
rather than $N$ that is fixed. The latter means that at $T=0$ there is
no phase separation. Instead,
$N(V_{g})$ dependence exhibits a jump from zero to a 
finite value of the interstitial concentration $N=N_{c}>0$. The
final state of the conducting channel in this case
is uniform. As it is usual for first order phase transitions,
the states with $N=0$ and $N=N_c$ are separated by an energy barrier. 
Thus, the
formation of the conducting channel is a collective phenomenon which 
involves 
tunneling of many electrons into the channel. 
Therefore, at this point, we can
not rule out a possibility that the $G(V_{g}, T,H)$ dependence is
hysteretic and 
that the experimental results depend on the history of the 
temperature,
the magnetic field and the gate voltage.

3). In the approximation of noninteracting
interstitials and at $\Delta_{\downarrow}<\epsilon_{0}$
 the band $k=1,\uparrow$ will be completely filled before the
the chemical potential reaches the top of the band $k=1,\downarrow$.
 This would lead to reentrance
from the plateau at $G=2\frac{e^2}{h}$ to the plateau at
$G=\frac{e^2}{h}$
as $N$
increases.

It is worth emphasising, however, that the presented above consideration 
is of a single particle nature. It neglects interstitial-interstitial
and interstitial-Wigner crystal lattice interactions and requires detailed
knowledge of interstitials band structures at relatively high interstitial
energies. In experimental situation,
$\frac{R}{a}=(\frac{\Delta_{\downarrow}}{J})^{1/4}$ is
not such a big number and the interaction effects could be very important
at large
$N$. In particular, the quasiparticle's bandwidth could depend on $N$. 

Some of these difficulties may be resolved
if we consider the opposite limit $\epsilon_{0}\ll \Delta_{\downarrow}$. In
this case
$k^{*}\sim \frac{\Delta_{\downarrow}-\Delta_{\uparrow}}{\epsilon_{0}}$
levels with spin
$\downarrow$ should be filled before the chemical potential of
 interstitials
reaches the band $k=1,\uparrow$.
Therefore the conductance is quantized in unit $\frac{e^{2}}{h}$ and
plateaus
in $G(V_{g})$ follow in a proper sequence. It is also
plausible that the
 first plateau
becomes
stronger as $T$ increases.
It is not clear, however, why in the framework of this picture the third
plateau at $G=\frac{3e^{2}}{h}$ should be more smeared than the second
one 
at $\frac{2e^{2}}{h}$.

In principle, it would be possible to distinguish  
$\epsilon_{0}>\Delta_{\downarrow}$ and $\epsilon_{0}<\Delta_{\downarrow}$
cases by different $V_{g}^{i}(H)$ dependences. Here
$V_{g}^{i}$
are the gate voltages at which the chemical potential reaches the bottom
of the
$i^{th}$ band and a new
plateau occurs. In the case
$\Delta_{\downarrow}<\epsilon_{0}$, the gate voltages $V_{g}^{1}$ and
$V_{g}^{2}$ correspond to bottoms
of
$k=1,\downarrow$ and $k=1,\uparrow$ bands. 
Therefore they move in
the opposite directions as $H$ increases. In the case
$\Delta_{\downarrow}>\epsilon_{0}$, gate voltages
$V_{g}^{1}$ and $V_{g}^{2}$ correspond to bottoms of $k=1\downarrow$ and
$k=2\downarrow$ bands and they move in
the same direction as $H$ increases.  
In experiments \cite{tom,tom1},  $V_{g}^{1}$ and $V_{g}^{2}$ move in
the
same direction. It is not clear, however, whether this can also be
explained as an
orbital magnetic field effect which exists due to a finite width of the
two
dimensional
electron gas \cite{tom,tom1}.

\section{conducting channels with large width}

In a uniform situation, the two-dimensional Wigner crystal melting is
believed
to
be a first order phase transition, which occurs at a critical electron
concentration $n_{c}$. In a nonuniform case shown in Fig.1,
a classical picture would predict that the transition takes place
at a larger concentration of electrons in the middle of the channel
$n(x=0)=n^{'}_{c}>n_{c}$
because of existence of the crystal-liquid surface energy.
The width of the conducting channel at the point of transition
is
$D_{c}=(\frac{3\alpha A^2}{2\gamma})^{1/3} \gg a$. Here
$\gamma=n_c \frac{\partial [E_{L}(n)-E_{W}(n)]}{\partial n}|_{n=n_{c}}$, 
$E_{L}(n)$ and $E_{W}(n)$ are 
energy densities of 2D Fermi liquid and Wigner crystal respectively, and
$\alpha$ is the surface energy per unit length.

Now consider a situation
when the width $D$ is large and the concentration of electrons in the
channel is high. 
In this case we can view the system as a stripe of Fermi
liquid  sandwiched between two antiferromagnetic (or RVB) Wigner crystal
(See Fig.3a).

It has been pointed out in the context of solid-liquid $^{3}He$
interface \cite{mey} that the
ferromagnetic ordering is
a generic property of a quantum rough surface.
It originates from the competition between kinetic
 energy of 
defects on the surface and the antiferromagnetic exchange energy in the
bulk of crystal. 
Since it is not known whether the Wigner crystal-Fermi liquid surface is 
in a
quantum smooth or in a quantum rough state, we
will consider the
case 
of a quantum smooth boundary, 
which is less favorable for the ferromagnetism, and present a model which
exhibits boundary
ferromagnetism even in this case.

We will model the system with the help of the Hubbard Hamiltonian defined
on
a
semi-infinite two-dimensional square lattice shown in Fig.3a. 
\begin{equation}
H_{D}=t\sum_{i,j,\sigma}
a^{\dagger}_{i\sigma}a_{j\sigma}+c.c.+U\sum_{i}
a^{\dagger}_{i\downarrow}a_{i\downarrow}a^{\dagger}_{i\uparrow}a_{i\uparrow}.
\end{equation}
Here $i,j$ labels adjacent sites on a square lattice,
$a_{i\sigma},a^{\dagger}_{i\sigma}$ are
annihilation
and creation electron operators on a site $i$, 
$\sigma$ is a spin index labeling 
$\uparrow$ and
$\downarrow$  directions of electron spins, $t$
and
$U$ are parameters 
which describe intersite tunneling and intrasite electron repulsion
respectively. In the case of half-filling and at $t\ll U$, the Hamiltonian
describes a Mott insulator, which is an analog of the Wigner crystal.
The spin ground state of the
system is antiferromagnetic (or RVB)
with exchange energy of order $J\sim \frac{t^{2}}{U}$ \cite{mot}.

We will describe the semi-infinite two-dimensional Fermi-liquid with 
the help of the Hamiltonian for noninteracting electrons
$H_{L}=\sum_{\bbox{q}}\epsilon_{\bbox{q}}b^{\dagger}_{\bbox{q}}
b_{\bbox{q}}$, 
where
$b_{\bbox{q}}$, $b^{\dagger}_{\bbox{q}}$ are annihilation and creation
electron operators in the liquid, 
$\epsilon_{\bbox{q}}=\frac{\bbox{q}^{2}}{2m}-E_{F}$ is the electron
energy, $\bbox{q}$ and $m$ are
the
momentum and mass respectively, $E_{F}$ is the liquid's Fermi energy.
The tunneling between the Mott insulator and the Fermi liquid is described 
by the Hamiltonian
\begin{equation}
H_{T}=t_{1}\sum_{i',\bbox{q}} a_{i}b_{\bbox{q}}^{\dagger}+cc.
\end{equation}
Here index $i'$ labels the localized states in the insulator at the
boundary.
The total Hamiltonian  is $H=H_{D}+H_{L}+H_{T}$.
In the case of half filling, the chemical potential of the Mott insulator
lies in the middle of the gap between the valence and the conduction band
(See
Fig.3b).
For a Wigner crystal, it means that 
the energy of creation of a vacancy and an interstitial electron would be
the same. In general, there is no reason to expect such a symmetry in a 
real 
Wigner crystal. Therefore we introduce into the model a small concentration of
impurity levels, which pin the position of the chemical potential
 below (or
above) the middle of the gap. These impurities are marked by
symbols "$+$" in Fig.3b. The distance between the impurities will be
much 
larger than all other characteristic lengths in the problem.
To be concrete we consider the situation when
the chemical potential is pinned near the bottom of the valence band
and
the energy for transferring an electron into the liquid and
creation of a vacancy in the Mott insulator is $U_{1}\ll U$ (See Fig.3b).

The tunneling between the liquid and the Mott insulator
leads to a negative correction to the ground state energy of the system.
The value of the correction depends on the spin
configuration in the Mott insulator. In the second order
perturbation theory with respect to $t_{1}$, we have an expression for the 
correction to the
energy per unit surface length 
\begin{equation}
\delta E^{t_{1}}=-\frac{a^3}{L_{x}L_{y}D}\sum_{{\bf q}, {\bf n}}
\frac{t^{2}_{1}}{U_1+\epsilon_{\{ S\}}
({\bf n})+\epsilon_{\bf q}}
\end{equation} 
where $a$ is the lattice spacing, $\epsilon_{\{ S\}}
({\bf n})$ is the energy of a vacancy moving in a spin
configuration $\{S\}$ in Mott
insulator, ${\bf n}$ labels the vacancy eigenfunctions in the Mott
insulator and $L_{y}$ is the sample length in {y} direction.
Suppose that in the Mott insulator near the surface there is
a ferromagnetic region of the width $L_{x}$ (See Fig.3a). Then we can
estimate
$\epsilon_{\{ S\}}
({\bf n})\sim \Delta cos(p_{y}a)+
k^{2}_{x}\frac{\hbar^{2}}{2m^{*}L_{x}^{2}}$. Here $\Delta =2t$ is the
vacancy energy bandwidth in a completely spin polarized Mott insulator,
$k_{x}$ is an integer corresponding
to quantized motion of the vacancy in $x$-direction, $p_{y}$ is the
quasi-momentum of the vacancy in   $"y"$ direction (${\bf
n}=k_{x},p_{y}$).
We assume that in the case of the antiferromagnetic ordering the
bandwidth of the vacancy is small and it is almost
localized.

Expanding Eq.6 with respect to  $\frac{t_{1}}{U_1}\ll 1$ and taking into account  the energy of
the spin polarization in the stripe of the width $L_{x}$, we get an
estimate for spin configuration
dependent part of the ground state energy 
\begin{equation}
\delta E_{\{S\}}\sim-\frac{t_{1}^{2}t^{2}}{E_{F}U_{1}^{2}L_{x}}+
\frac{JL_{x}}{a^{2}}.
\end{equation}
Eq.7 has a minimum at 
\begin{equation}
L_{x}=a\sqrt{\frac{t_{1}^{2}U}{U_{1}^{2}E_{F}}}.
\end{equation}
Therefore the system is magnetically polarized at the surface provided 
$\frac{t_{1}^{2}U}{U_{1}^{2}E_{F}}\gg 1$.

The statement about existence of the magnetic polarization near the surface 
can be proved in a more rigorous way starting with usual 
perturbation theory
\begin{equation}
\Delta E^{t_{1}}= \frac{t^{2}_{1}}{\epsilon_F}\sum_{i,i',\Gamma(i,i')}
\int d\epsilon
\frac{C(n_{\Gamma})}{\epsilon + U_1}
(\frac{t}{\epsilon + U_1})^{n_{\Gamma}}  J_1(i,i').
\end{equation}
Here $i,i'$ labels the lattice sites at the boundary,
$\Gamma(i,i')$
labels paths of a vacancy on the lattice intersecting the boundary at
sites $i$ and $i'$,
which preserve an initial spin configuration of the system ( An example of
such a path is shown
in Fig.3a), $J_{1}(i,i')$
is a function of the distance between sites $i$ and $i'$, which reflects
the Friedel oscillations in the Fermi liquid,
$n_{\Gamma}$
is the number of steps a vacancy hops along a path $\Gamma$.
While obtaining Eq.9 from Eqs.4,5, we neglected all terms which involve
double electron occupancies of lattice sites.
 The factor
$C(n_{\Gamma})$ depends on the spin configuration in the system. 
In the ferromagnetic case, due to Fermi statistics of electrons, it is
equal to $(-1)^{n_{\Gamma}}$.
In the limit $p_F a \gg 1$, $J_1 =\delta_{i,i'}$ and we only need
to take into account paths which enter and leave the Mott insulator
through the same lattice site $i=i'$. 
At this point the problem is reduced to that considered in
\cite{nag,thouless}, where it
was proven that the ground state of the system with one vacancy and  infinite $U$ is
ferromagnetic. A resummation of these ferromagnetic paths
leads
to Eq.6.

If the surface is in a quantum
rough state
the ferromagnetic ordering at the surface is even more favorable
energetically \cite{mey}.

The presented above picture is relevant when the width $D$ of the Fermi liquid
stripe is large, which corresponds to large $k_{1}$ in Eq.1.
Strictly speaking, in this case, each plateau with given $k_{1}$ also
splits
into two plateaus quantized in unit $\frac{e^{2}}{h}$. However, the
amplitude
of this splitting of plateaus should be small.

Let us consider what happens as concentration of electrons in
 the channel and, consequently, $D$ decreases.
It has been mentioned above that the mean field theory, which neglects quantum
fluctuations 
of the position of the crystal-Fermi liquid boundary, predicts the creation of
the conducting channel
as a first order phase transition.
The question arises whether quantum fluctuations of electron positions in the
Wigner crystal can
make this transition to be a second order one.
 The amplitude of quantum fluctuations of the position of the
 crystal-liquid boundary 
depends on whether the surface is in a quantum rough or a quantum smooth
state.
 If it is in a quantum rough state, we can describe deviations of the
boundary
from its
equilibrium classical
 position by a variable $X(y,t)$ and use the
following effective action
\begin{equation}
S=\int dy dt[\beta(\frac{dX}{dt})^{2}+\alpha(\frac{dX}{dy})^{2}+\tilde{V}(X)]
\end{equation}
to estimate the amplitude of quantum fluctuations of the position of the
boundary.
Here $\tilde{V}\sim \xi X^{2}$ corresponds to the energy difference
between
the solid and
liquid phase as the boundary is moved by distance $X$,
$\xi=\gamma\frac{D_{c}}{A^{2}}$. The second term in
Eq.10 corresponds to the surface tension and the first term 
corresponds to acceleration of steps along the surface.

We would like to mention that the situation described by Eq.10 is 
different
from that considered in
\cite{andpa}. 
The electron density profile $n(x)$ in the channel is
determined by
the long range Coulomb force.
Therefore, in our case there is no redistribution of the electron 
density in
the liquid associated  with oscillations $X(y,t)$.
Using Eq.10 we get an
estimate 
 $<(\delta X)^{2})>\sim
\frac{1}{\sqrt{\alpha\beta}}\log{\frac{\alpha}{\xi a}} $. Here the
brackets $< >$ stand for the quantum mechanical
averaging.
If $<(\delta X)^{2})>\gg D_{c}^{2}$, then
 the
one-dimensional
 melting can occur 
as a second order phase transition and we arrive at the picture considered
in 
the previous section.
 We
believe that, in principle, this inequality can
take place because the transition occurs at
$r_{s}\sim 39$ where at short distances the crystal differs
 from the liquid very little.

In conclusion we would like to make the following remarks.
At finite temperature the ferromagnetic ordering of the quasi-one
dimensional stripe will  be destroyed by thermal fluctuations.
Strictly speaking, at $T=0$, the presented above arguments only prove
existence
of strong ferromagnetic correlations at "small" distances. At this stage
we can not rule out a possibility that at "large" distances the
ferromagnetism will be destroyed by quantum fluctuations of spin
configuration in a 2-D Wigner crystal. Quantization of the conductance of
the channel in units 
$\frac{e^{2}}{h}$ , however, persists as long as the length of
the channel is shorter than the characteristic distance of the destruction
of the ferromagnetism.

It has been shown \cite{1D} that the
ground state of strictly one-dimensional systems can not be ferromagnetic.
However, in the
experiment
\cite{tom,tom1} the system is not strictly one-dimensional. We are 
unaware of any
theorem which prevents existence of ferromagnetism in quasi-one-dimensional
system of the type
shown in Fig.3. We leave this questions for future investigation.

We are grateful to C. Markus,
D. Thouless, S. Kivelson, D. Khmelnitskii, E. Lieb and M. den Nijs for
useful discussions.   
Work of B.S. is supported by Division of Material Sciences, U.S.National
Science Foundation under Contract No.DMR-9625370.
F.Z. is supported by ARO under DAAG 55-98-1-0270.
F.Z. also likes to thank NECI in Princeton for its hospitality.

\newpage

\begin{figure}
  \centerline{\epsfxsize=12cm \epsfbox{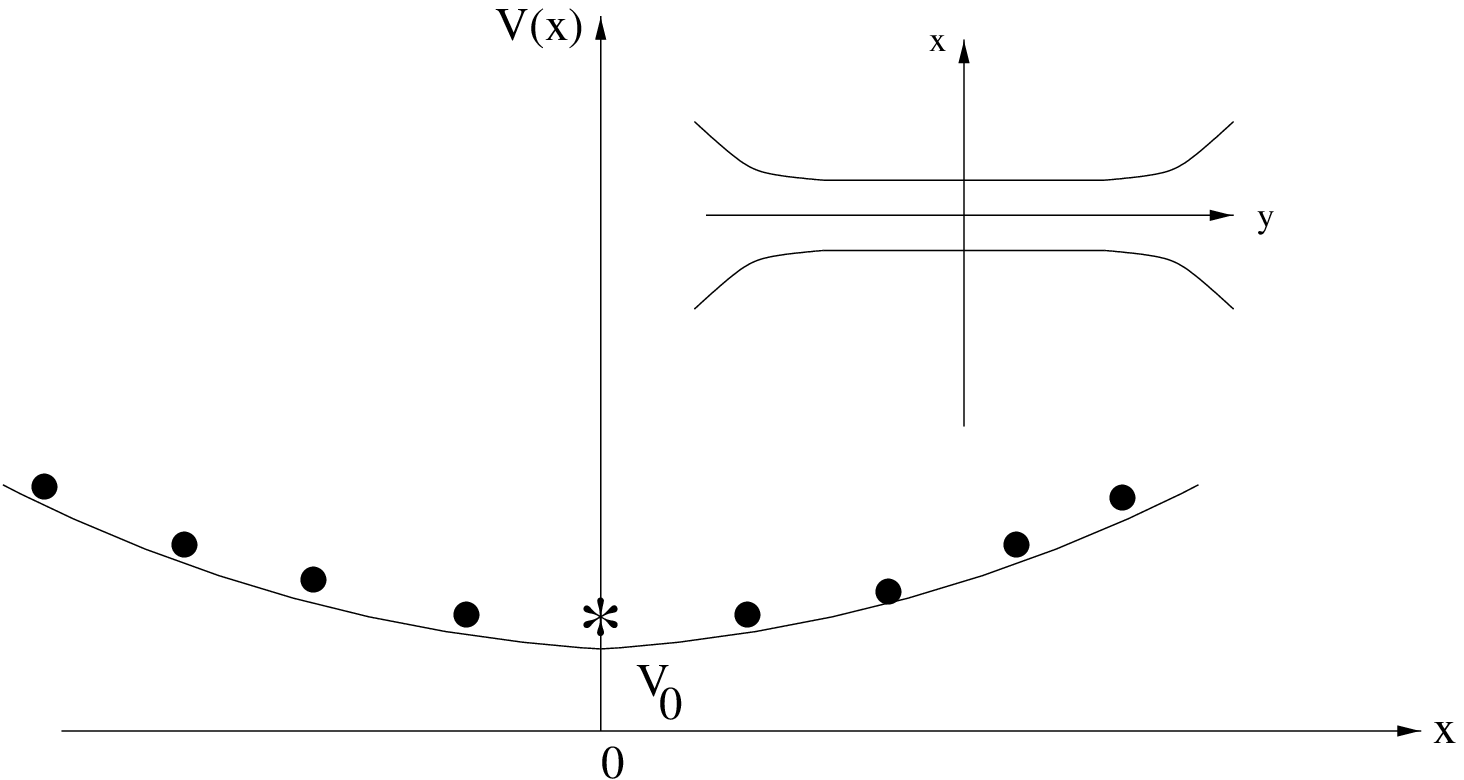}}
\vspace{1.5cm}
  \caption{ 
The qualitative picture of the confining potential $V(x)$. Solid 
dots correspond to positions of electrons in the Wigner crystal. The symbol
$"*"$ corresponds to an interstitial electron. The insert shows the
geometry of the conducting channel.
} \
\label{fig:fig1}
\end{figure}

\newpage

\begin{figure}
  \centerline{\epsfxsize=10cm \epsfbox{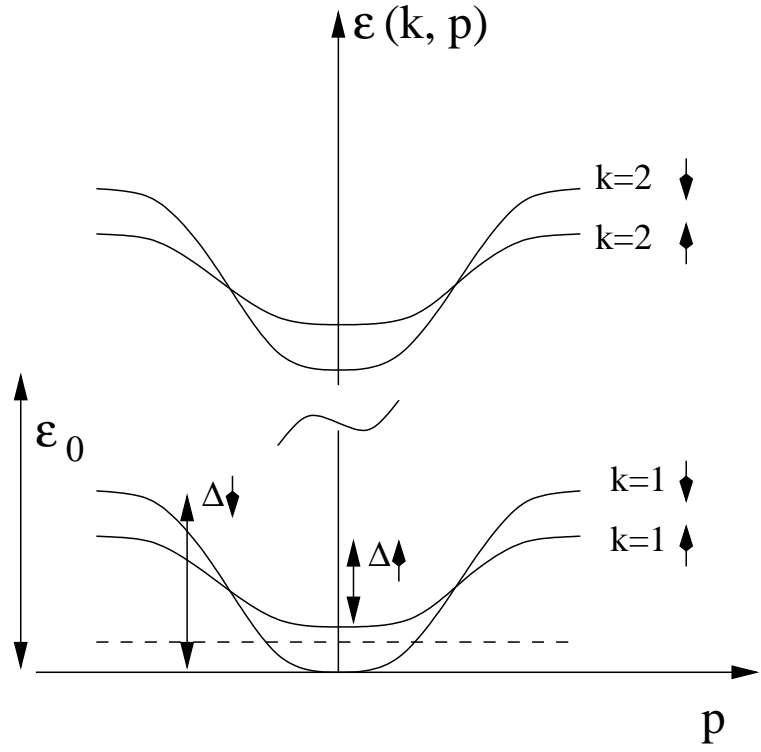}}
  \vspace{1.5cm}
\caption{
A schematical picture of band structures of interstitial
electrons in the case
$\epsilon_{0}\gg \Delta_{\downarrow}$.
The dashed line indicates the chemical potential when the channel 
becomes conducting.
}\

  \label{fig:fig2}
\end{figure}

\newpage

\begin{figure}
  \centerline{\epsfxsize=14cm \epsfbox{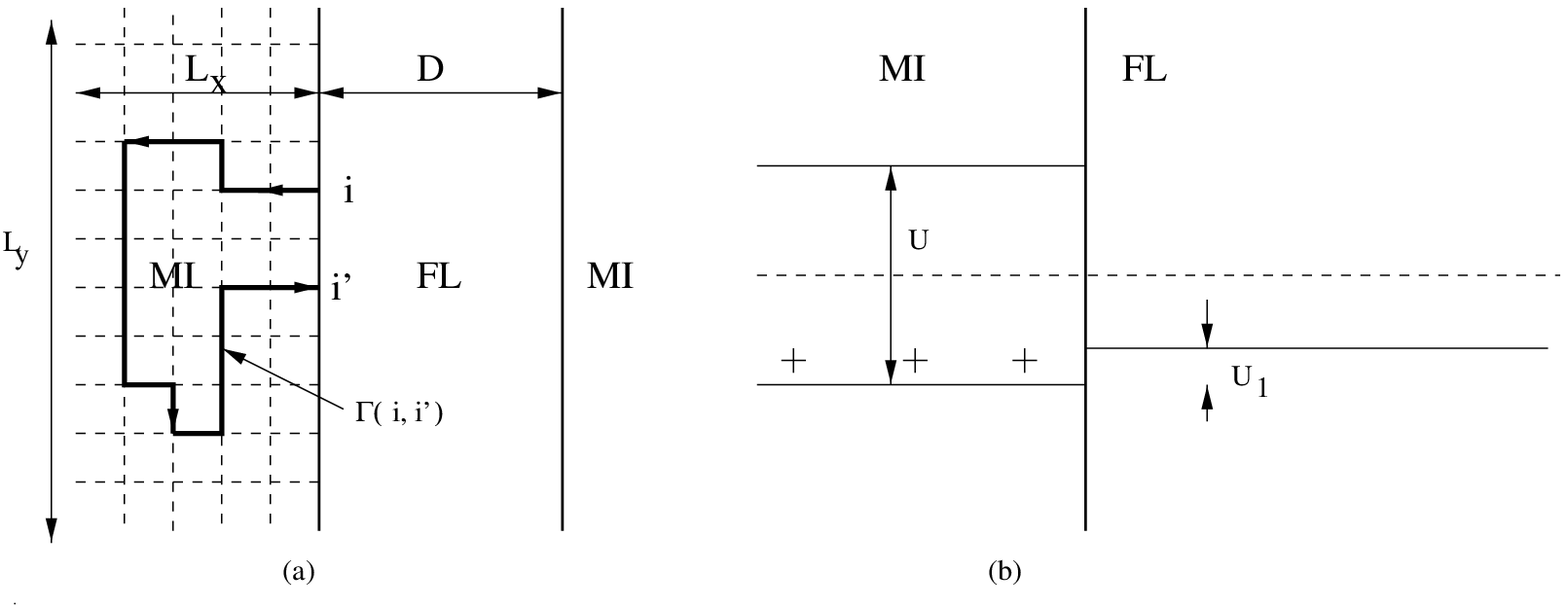}}
\vspace{1.5cm}  

\caption{
a. The geometry of the Mott insulator (MI)-Fermi liquid (FL)-Mott
insulator structure. The Mott insulator and the Fermi liquid are indicated
by symbols "MI" and "FL" respectively.
$\Gamma$ represents a path corresponding to a term in the
perturbation sequence in Eq.9. 
b. Energy diagram for the Mott insulator-Fermi liquid 
boundary. 
}
\
  \label{fig:fig3}
\end{figure}


\begin{thebibliography}{99}

\bibitem{tren}T. J. Thornton et.al.,  Phys. Rev. Lett. {\bf 56},
1198(1986).
\bibitem{wees}B. J. van Wees, et.al., Phys. Rev. Lett. {\bf 60},
848(1988).
\bibitem{tom}
K. J. Thomas, J. T. Nicholls, M. Y. Simmons,
M. Pepper, D. R. Mace and D. A. Ritchie, Phys. Rev. Lett. {\bf 77},
135(1996);
\bibitem{tom1} K. J. Thomas, J.T.Nicholls, M.J.Appeyard, M.Y.Simmons,
M.Pepper, D.R.Mace, W.R.Tribe, D.A. Ritchie, Phys.Rev. {\bf B58},
4846(1998).
\bibitem{binakk}C.W.J.Beenakker, H.van Houten, in 
{\it Solid State Physics}, edited by
H.Ehrenreich and D.Turnbull (Academic Press, NY, 1991).
\bibitem{andlif}A. Andreev, I. Lifshitz, Sov.Phys. JETP {\bf 29}, 
1107(1969)[Zh. Eksp. Teor. Fiz. {\bf 56}, 2057(1969)].
\bibitem{andpa}A. F. Andreev, A. Ya. Parshin, Sov. Phys. JETP {\bf 48},
763(1978)[Zh. Eksp. Teor. Fiz. {\bf 75}, 1511(1978)].
\bibitem{paba}K. O. Keshishev, A. Ya. Parshin and
A. V. Babkin, Sov. Phys. JETP {\bf 53}, 500(1981)
[Zh. Eksp. Teor. Fiz. {\bf 80}, 716(1981)].
\bibitem{fish}D. S. Fisher, J. Weeks, Phys. Rev. Lett. {\bf 50}, 1077
(1983).
\bibitem{ior}S.V.Iordansky, S.E.Korshunov, J.Low.Temp.Phys. {\bf 58}
425(1985).
\bibitem{chak}S. Chakrovarty, S. Kivelson, C. Nayak and K. Voelker,
cond-mat/9805383.
\bibitem{her}C. Herring, Rev. Mod. Phys. {\bf 34}, 631(1962).
\bibitem{ferow}M. Roger, Phys. Rev. {\bf B 30}, 6432(1984).
\bibitem{rvb}P.W. Anderson, Mater. Res. Bull.,{\bf 8}, 153(1973).
\bibitem{nag}Y. Nagaoka, Phys.Rev.{\bf 147}, 392 (1966);
Solid State Commun. {\bf 3}, 409(1965).
\bibitem{thouless}D. J. Thouless, Proc. Phys. Soc. (London) {\bf 86},
893(1965).
\bibitem{nagae}E. L. Nagaev, Sov. Phys. JETP {\bf 27}, 122(1968) 
[Zh. Eksp. Teor. Fiz. {\bf 54}, 228(1968)].
\bibitem{nagae1} E.L.Nagaev, "Physics of magnetic semiconductors", Nauka, 
Moskow, (1979).

\bibitem{brink}W. F. Brinkman, T. M. Rice, Phys. Rev. {\bf B2},1324(1970).

\bibitem{and}A. F. Andreev, JETP Lett.{\bf 24}, 564(1976)[Pisma Zh. Eksp. 
Teor. Fiz. {\bf 24}, 608(1976)]. 

\bibitem{spiv}B.Spivak, Sov. Phys. JETP {\bf 60}, 787 (1984).


\bibitem{kris} A. Kristensen, et. al., cond-mat/9808007.

\bibitem{kris1} A.Kristensen,  et. al., Physica {\bf B}: Condensed Matter
{\bf 249-251}, 180 (1998).

\bibitem{kiv}S. A. Kivelson, E. Fradkin and V. J. Emery, {\em Nature}
{\bf 393}, 550(1998).

\bibitem{mey}M. A. Meierovich, B. Z. Spivak, JETP Lett. {\bf 34},
551(1981)[Pisma Zh. Eksp. Teor. Fiz. {\bf 34}, 575(1981)]

\bibitem{mot}N. F. Mott, {\it Metal-insulator transitions}, Taylor and
Francis, London(1974).


\bibitem{1D} E. Lieb, D. Mattis, Phys. Rev. {\bf B125}, 164(1962).
\end{thebibliography}
\end{document}